# Nitrogen doped $In_2O_3$-ZnO nanocomposite thin film based sensitive and selective ethanol sensor

P. K. Shihabudeen[1] and Ayan Roy Chaudhuri[1]*

[1] *Materials Science Centre, Indian Institute of Technology Kharagpur, 721302 Kharagpur, West Bengal, India*

**Abstract**

Nanocomposite metal oxide thin films exhibit assuring qualities in the field of gas sensors because of the collective opportunities provided by the heterointerface formation. In this work, we present the synthesis of nitrogen doped mesoporous $In_2O_3$-ZnO nano-composite thin films by simple wet chemical method using urea as the nitrogen precursor. SEM investigation suggests formation of mesoporous nano-composite thin films, where the uniformity of surface pore distribution depends on the relative proportion of $In_2O_3$ and ZnO in the composites. HRTEM investigation suggest formation of sharp interfaces between N-$In_2O_3$ and N-ZnO grains in the nano-composite thin films. The nano-composite thin films have been tested for their ethanol sensing performance over an extensive range of temperature, ethanol vapor concentration and relative humidity. Nitrogen doped nano-composite thin film with equal proportion of $In_2O_3$ and ZnO exhibits excellent ethanol sensing performance at a reasonable operating temperature (~94 % at 200 °C for 50 ppm of ethanol), fast response time (~two seconds), stability over time, enhanced resilience against humidity and selectivity to ethanol over various other volatile organic compounds. All the

---

* Corresponding author: Tel: +91-3222-283978

Electronic mail: ayan@matsc.iitkgp.ac.in (Ayan Roy Chaudhuri)

results indicated that nitrogen doped $In_2O_3$/ZnO nano-composite thin film portrays great possibilities in designing improved performance ethanol sensor.

*Keywords:* Zinc oxide, Indium oxide, heterostructure, ethanol detection

**1. Introduction**

Ethanol is a volatile organic compound (VOC) commonly used in industrial as well as domestic applications [1,2]. Increased utilization of ethanol in industry rises the danger of explosion as well as ground-water contamination [3,4]. Moreover, the leading source for road accidents around the world is due to driving under the influence of alcohol. In addition, metabolically generated ethanol in breath also acts as biomarker related to liver diseases such as cirrhosis and fatty liver [5]. Semiconducting metal oxide (SMO) based thin film gas sensor is a promising candidate to monitor VOCs offers advantages like chemical stability, good sensitivity, detection of wide range of gases etc. [6,7]. Despite their promising sensing performance toward VOCs the SMO based gas sensors suffer from drawbacks such as high operating temperature, slow sensing and cross-sensitivity [8,9]. In order to circumvent the issues, numerous approaches have been utilized like surface microstructure modification, multiple metal oxide nanocomposite thin films, composites with noble metals, dopant incorporation etc [10,11].

Among these approaches oxide nanocomposites and incorporation of suitable dopants in oxides offers distinct advantages. Fabrication of nanocomposite thin films significantly enhances the sensing characteristics due to their varying compositions and the synergistic effects (geometrical effects, chemical effects and electronic effects) in the composites [12,13]. The nanoscale heterojunctions in the composite thin films modify the electronic structure and increases the oxygen vacancy concentration, which in turn affect the selectivity and sensitivity towards gases [14,15]. $In_2O_3$ and ZnO are two common SMOs which have been extensively studied for gas sensing. The composites and heterostructure of these two oxides in form of like microsphere [16], nanosheet [15], nanofiber [17] and nanowire [18] are reported for ethanol sensing application. Even though the hetero structures have displayed superior ethanol sensing performance compared to the individual oxides, they still suffer from high operating temperature, slow response time and cross-sensitivity [14,19]. Additionally, the complex nanostructures often suffer from temporal and thermal instabilities. In this context thin film-based sensors are considered as suitable alternative to overcome the shortcomings of nanostructures [20].

In contrast, doping induces changes in electronic structure, chemical bonding and oxygen vacancy concentration which leads to selective adsorption of gas molecules [21,22]. In our previous reports on nitrogen doped $In_2O_3$ and ZnO thin films, an improved performance on ethanol sensing characteristics compared to their corresponding pristine oxides has been observed [22]. The N-$In_2O_3$ thin film has displayed a fast and stable response, whereas N-ZnO thin film exhibited selective and low temperature sensing. In both cases changes in band gap and increase in oxygen vacancy has been observed. Based on our observations it appeared that nano composite thin film-based sensors consisting of N-$In_2O_3$ and N-ZnO shall be an efficient approach of producing fast and selective chemiresistive ethanol sensor.

This work focuses on preparation of nanocomposite thin films of N-$In_2O_3$/N-ZnO with different concentration of the nitrogen doped oxides in the composite and elucidate their ethanol sensing characteristics. We demonstrate that the nitrogen doped composite thin film with optimum composition exhibit superior sensing response, faster response time, selectivity towards ethanol, and excellent stability over a wide range of relative humidity condition in comparison to the undoped composite sensor as well as nitrogen doped individual oxides.

2. Experimental

2.1 Thin film synthesis

Urea (Alpha-aeser, >99 %), Zinc acetate dihydrate (Merk, >99%) and Indium acetate (Alfa-aeser, >99%) were used as precursor for nitrogen, zinc and indium respectively for the thin film synthesis. At first, the sol has been synthesised, for which a solution of zinc acetate and indium acetate was prepared in 10 ml of ethanol (99.9 %) as shown in table S1. Ethanolamine (3ml), was added drop wise to the solution while stirring at 60 °C. Urea (1.2 g) was dissolved in 10 ml of ethanol in a separate beaker at similar stirring conditions. Thereafter, these two solutions were mixed together and it was stirred at 60 °C for 4 hours till a stable reddish-orange solution was obtained [22]. Thin films have been prepared by spin coating of the sol at 3000 rpm for 20 seconds on pre-cleaned $SiO_2$/Si substrates. After coating, the solvents were removed by drying it on a hot plate at 125 °C. The composite film with similar thickness was prepared by coating the sol multiple times. Finally the samples have been annealed at 750 °C for one hour to obtain crystalline thin films. Samples were named IZON 25, IZON 50 and IZON 75 which has 25 %, 50 % and 75 % $In_2O_3$ in the composition respectively. Similarly, a reference sample IZO 50 without urea was synthesised following above-mentioned procedure.

2.2 Material characterisation

Panalytical Empyrion X-ray diffraction (XRD) Machine with a Cu K-α radiation source (1.54 Å) was used to determine the crystal structure of the synthesised thin films. Field emission scanning electron microscope (FESEM Zeiss Gemini, Germany) was used to analyse the surface microstructure and cross-section of the samples. The transmission electron microscopy analysis was performed using FEI TECNAI G2 F20 X-TWIN. X-ray photoelectron spectroscopy (model PHI 5000 Versa Probe II, INC, Japan) with A1 K-α x-ray source (1486.6 eV) was used to study the surface chemical composition.

2.3 Sensor measurements

Gold was used as the electrodes for ethanol sensing characteristics. An interdigitated electrode with 2 mm width and 1.5 mm electrode separation was sputtered on to the thin films. An inhouse built quasi-static gas sensing setup was used to carry out the VOC sensing measurement. A load resistance ($R_L$), which has a similar resistance of the sensing film is connected in series and a continuous ~5 V DC is supplied over the sensor. A data collecting arrangement based on a microcontroller (Atmel ATMEGA 32) is employed to measure the net output voltage across $R_L$. For additional evaluation and data storage, the whole setup is connected to a desktop computer via the RS232 interface. The detailed measurement process could be found in our previous work[22]. The sensor response (S) of the thin films were calculated using the following relation

$$S(\%) = \frac{(R_a - R_g)}{R_a} \times 100 \quad (1)$$

Where $R_g$ and $R_a$ are the equilibrium resistances measured in the presence of test gas and air, respectively. The response ($\tau_{res}$) and recovery ($\tau_{rec}$) time were estimated as the time taken to raise base resistance to 90% of the maximum resistance and the time taken for resistance to fall 90% of the equilibrium resistance in the air.

## 3. Material characterization

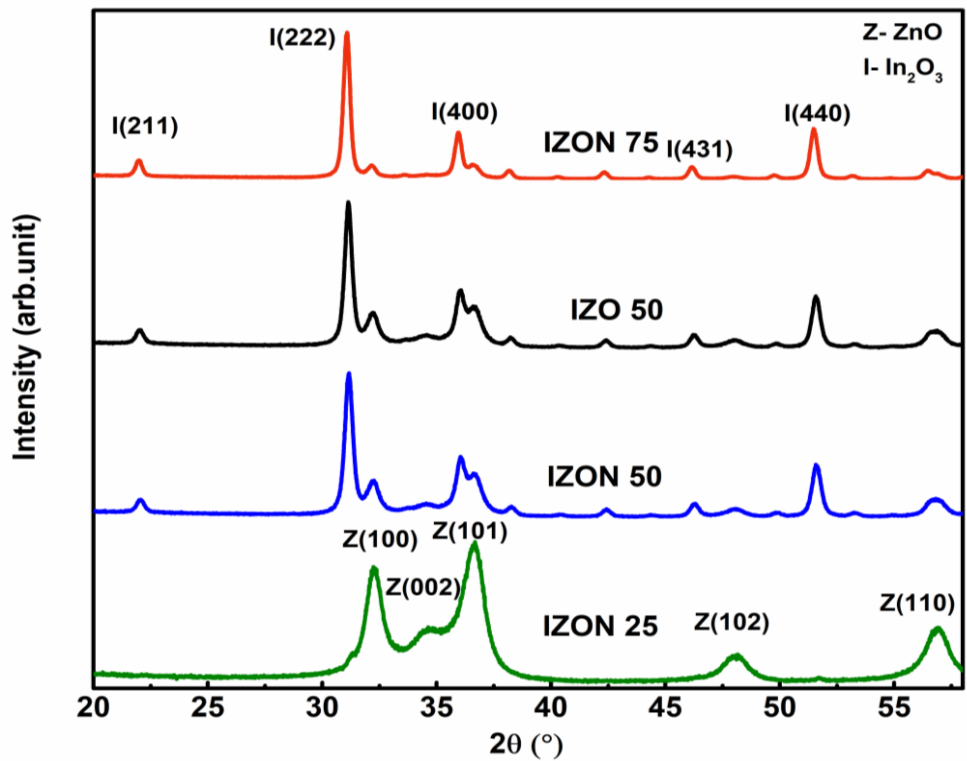

Figure 1. XRD of various composite thin films.

Figure 1 portrays the XRD patterns of different composite thin films. The diffraction peaks have been marked in accordance to typical diffraction pattern for cubic $In_2O_3$ (JCPDS 006-0416) and wurtzite ZnO (JCPDS 36-1451). All the films are well crystalline which exhibit diffraction peaks corresponding to both $In_2O_3$ and ZnO. While ZnO rich (75%) IZON 25 has shown intense peaks which correspond to ZnO (100),(002) and (101) planes, with increase of $In_2O_3$ fraction, intensity of the $In_2O_3$ peaks increase. No obvious shift has been observed with nitrogen doping for composite films of similar composition, e.g. IZON 50 and IZO 50. On the other hand a slight shift in peaks has been observed with variation in film composition (figure S1).

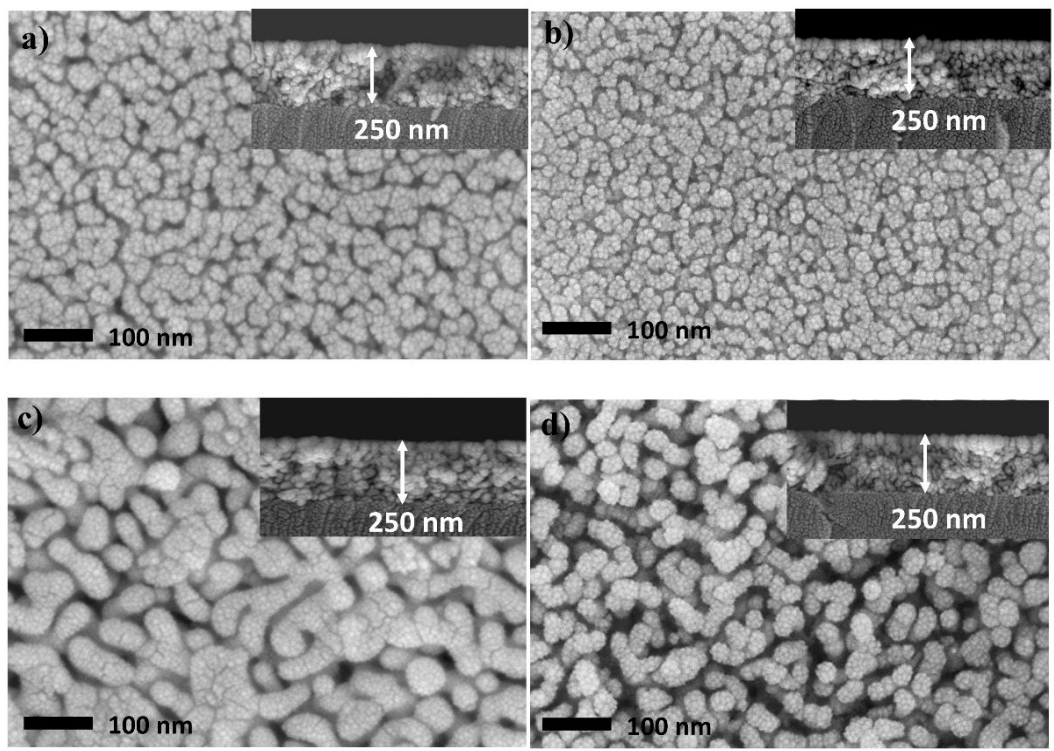

Figure 2. scanning electron microscopy images of a) IZON 25, b) IZON 50, c) IZON 75 and d) IZO 50 composite thin films (cross-sectional SEM in the inset).

Scanning electron microscopy has been carried out to investigate the surface microstructure of composite thin films. Figure 2 depicts the SEM surface images of the thin films. All the films exhibit mesoporous surface morphology. The pore size distribution in the composite thin films has been estimated from the SEM image analysis (figure S2). While the IZON 50 thin film possesses an average pore diameter of ~24 nm, the IZO 50 film has larger pores with average diameter of ~ 52 nm. IZON 25 thin film has been found to possess pore size (~26 nm) similart to that of IZON 50, but with larger agglomerated grains. On the other hand, IZON 75 formed larger pores (~41 nm) and agglomerated particles. Large amount of $In_2O_3$ in IZON 75 and ZnO in IZON 25 leads to an agglomerated structure and reduces surface pore disribution on the the film [23]. The surface pore coverage for the composite thin films has been estimated from the post image analysis using ImageJ software of corresponding SEM images (figure S3). The estimated values are 30%, 40 %, 22 % and 26 % for IZON 25, IZON 50, IZON 75 and IZO 50 thin films respectivly. All the samples have comparable thickness of ~250 nm as observed from the cross-sectional SEM images (inset, figure 2).

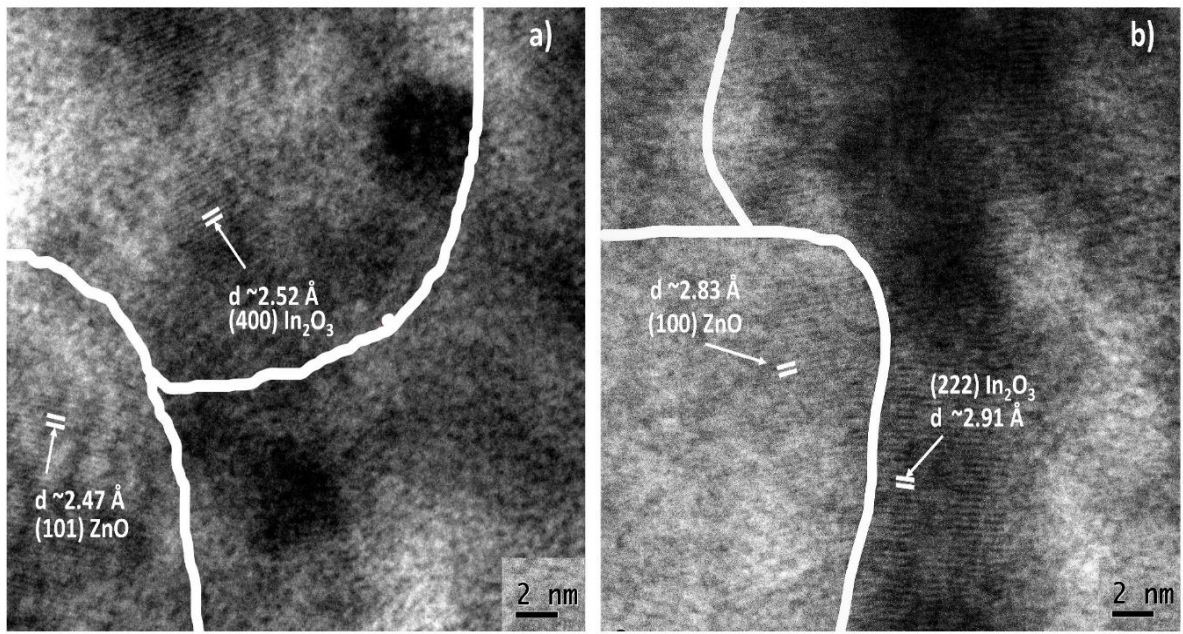

Figure 3. TEM images of a) IZON 50, b) IZO 50 composite thin films.

In order to confirm the formation of heterojunciton between the two oxides in the nanocomposite thin films, the samples have been investigated by transmission electron microscpy (TEM). Figure 3 displays TEM images of IZO 50 and IZON 50 as representative samples. The TEM images confirm high crystallinity of cubic $In_2O_3$ and hexagonal ZnO. In both the composite thin films formation of nano-scale heterojunction without any detectable intermixing has been confirmed from figure 3(a-b), which display two distinct regions (marked with white lines) with d spacing values corresponding to cubic $In_2O_3$ crystal structures and wurtzite ZnO. Creation of well-constructed heterojunction between ZnO and $In_2O_3$ with good crystallinity is important for the efficient interfacial charge carrier separation or transportation [24].

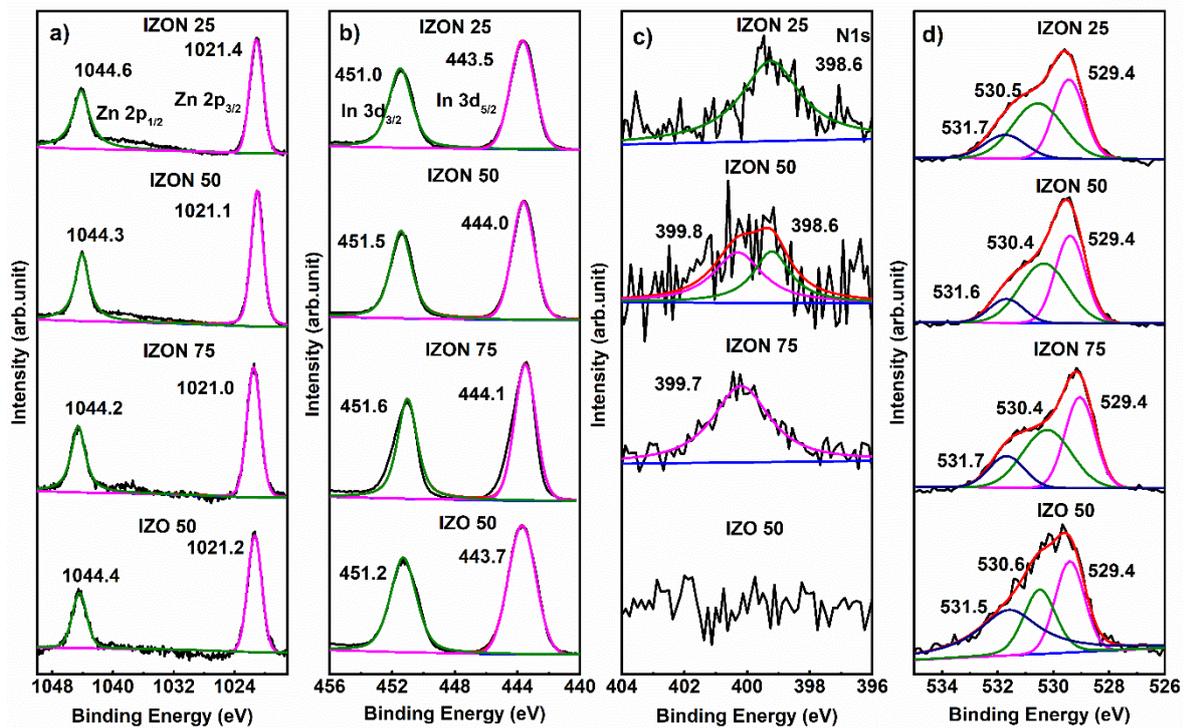

Figure 4. High resolution XPS spectrum of a) Zn 2p, b) In 3d, c) N 1s and d) O 1s of various composite thin films.

XPS analyses have been performed to evaluate the chemical composition and to confirm nitrogen doping in the composite thin films. Carbon 1 s peak (284.5 eV) has been used as a reference for calibration and Tougaar background have been used for the peak background correction [25]. Figure 4 (a) and 4(b) display the Zn 3p and In 3d peaks of the composite thin films. The doublet peaks of In 3d and Zn 2p have been attributed to $In^{3+}$ and Zn $^{2+}$ states respectivly [16]. In 3d peaks exhibit a rigid shift to higher binding energy with increase in $In_2O_3$ composition in the composites, whereas, the Zn 2p peaks shifted to lower binding energy. The opposite shift of $In^{3+}$ and $Zn^{2+}$ peaks has been attributrd to transfer of electrons from low work function $In_2O_3$ to high work function ZnO across the nanoscale heterojunctions in the composites [24,26]. The difference in the binding energy positions of the $In^{3+}$ and $Zn^{2+}$ peaks in IZO50 and IZON50 can be attributed to the increase in n-type characteristics due to nitrogen doping [22,25]. In:Zn ratio for each composite films has been estimated as 28:72, 51:49, 76:24 and 51:49 for IZON 25, IZON 50, IZON 75 and IZO 50 respectivly. The composition of various composite films has been shown in table S2.

Figure 4(c) depicts the high-resolution XPS spectra of the N1s region (396-404 eV) for the composite thin films. While the undoped IZO 50 film does not exhibit any identifiable peak in this region, IZON 50 exhibits a broad peak, which has been deconvoluted into two peaks around 398.6 eV and 399.8 eV. The lower energy peak has been attributed to substitutional($N_s$) doping of N in ZnO and the other peak has been identified as interstitial($N_i$) doping of N in $In_2O_3$ in agreement to the previous literature reports[22,27]. The ratio of $N_i:N_s$ has been estimated to be 52:48 which is comparable to the estimated In:Zn ratio in the IZON 50 composite thin film. In case of IZON 25 and IZON 75, single N1s peaks have been observed around 398.6 eV and 399.7 eV respectivly which correspond to $N_s$ in ZnO and $N_i$ in $In_2O_3$ respectively [22]. Due to lower concentraion of N-$In_2O_3$ in IZON 25 and and N- ZnO in IZON75 their corresponding N 1S peaks (i.e. $N_i$ and $N_s$ respectively) could not be resolved. However, since all the samples have been prepared under identical conditions, it is envisaged that the $N_i:N_s$ would be similar to the In:Zn ratio in all the composite thin films. The nitrogen concentration for all the N-incorporated composite thin films has been estimated to be ~0.6 at%.

Figure 4(d) represents the O 1s spectra of composite films. For all the films, the O1s peak has been deconvoluted into three peaks corresponding to Metal-Oxygen (∼529.4 eV), non-stoichiometric oxygen (∼530.4 eV–530.6 eV) and surface adsorbed hydroxyl species (∼531.5 eV–531.7 eV) [16,28]. The peak corresponding to adsorbed surface hydroxyl has been more pronounced for IZO 50 film compared to the IZON films, which is attributed to reduced adsorption of water molecules on the nitrogen doped metal oxides [29,30]. The ratios of areas under curves $A_V/A_L$ ($A_V$ and $A_L$ are the area under the curve for non-stoichiometric and lattice oxygen respectively) for IZON 25, IZON 50, IZON 75 and IZO 50 have been estimated to be 1.12, 1.17, 1.08 and 0.85 respectivly. The increase in oxygen vacancy (related to non-stoichiometric oxygen) in the IZON films have been accredited to doping [31].

## 4. Ethanol sensing characteristics

Figure 5(a) compares the temperature-dependent sensing responses (%) of the IZON thin films with IZO 50 for 50 ppm of ethanol in the temperature range of 150 °C - 300 °C. All the sensing responses have been measured within an average error limit of ±0.3 %. IZON 25 and IZON 50 exhibited superior ethanol sensing response compared to IZO 50 at operating temperatures below 275 °C, while IZON 75 exhibited inferior response. At operating temperatures below 250 °C, IZON 50 exhibited the best sensing response against ethanol among all the composite thin films. In the present study the optimum working temperature has been chosen to be 200 °C at which IZON 50 exhibited a sensing response of ~ 94%. Figure 5(b) and (c) portray response and recovery transient of the thin film sensors at 200 °C against 50 ppm of ethanol. Response time of the IZON 50 sensor has been found to be the fastest (~2 seconds), which is followed by IZON 25, IZO 50 and IZON 75. The recovery time for all the IZON sensors have been found to be similar (~300 s), whereas IZO50 exhibited slightly lower recovery time (~ 250 s). Table S3 summarizes the response and recovery time of all the sensors against 50 ppm ethanol measured at 200 °C.

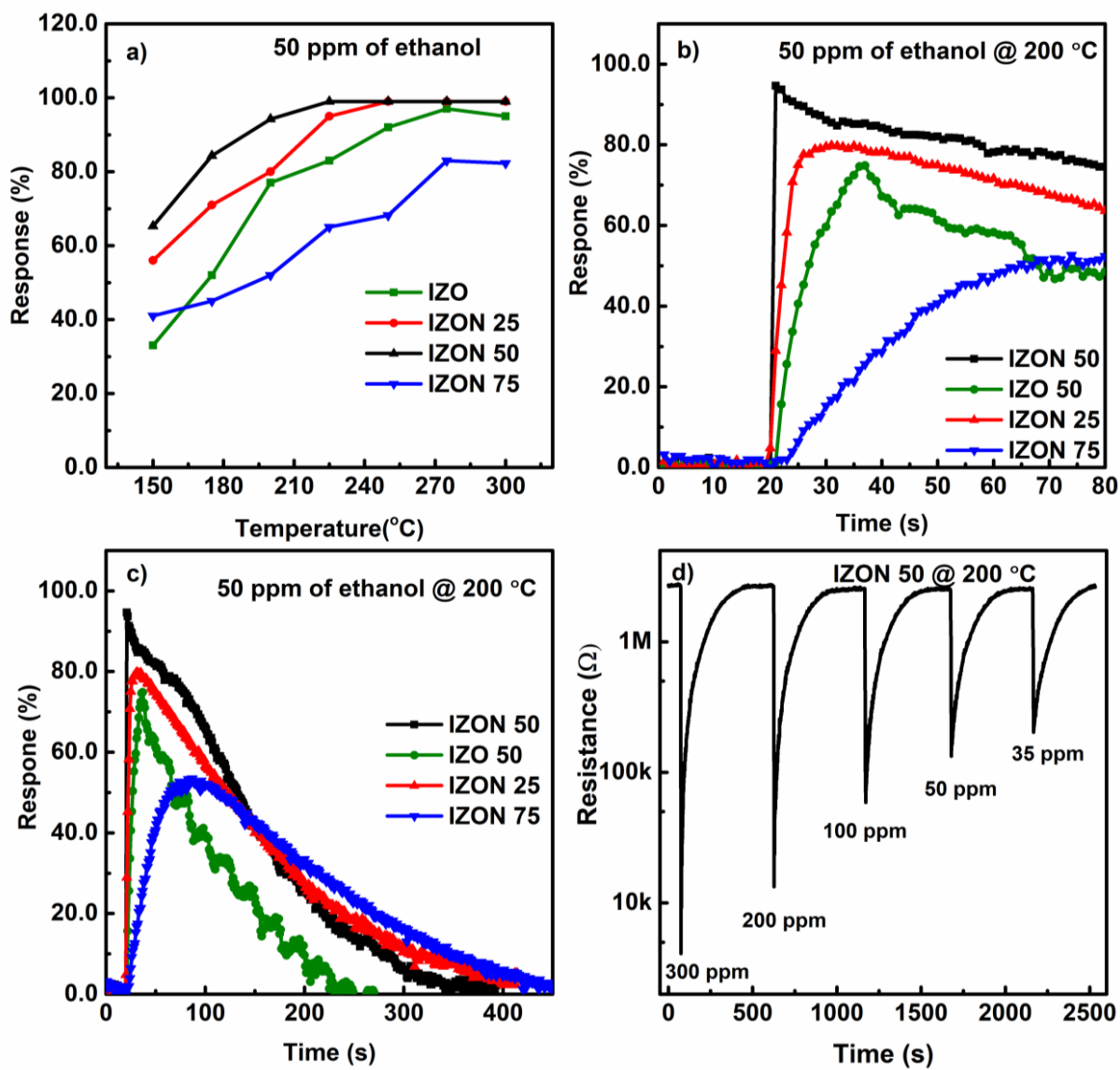

Figure 5 a)) ethanol sensing characteristics of various composite films over a range of temperatures, b) response, c) recovery transient of the various composite thin films 200 °C and d) concentration variation study of IZON 50 thin films at 200 °C.

Based on the superior performance of IZON 50, the sample has been chosen for further investigation. Figure 5(d) depicts the resistance transient of the IZON 50 composite thin film measured at 200 °C against varied ethanol concentration (35 ppm - 300 ppm). The sensor exhibited significantly higher response of ~99 % against 300 ppm ethanol which has changed to only ~92 % at ethanol concentration as low as 35 ppm.

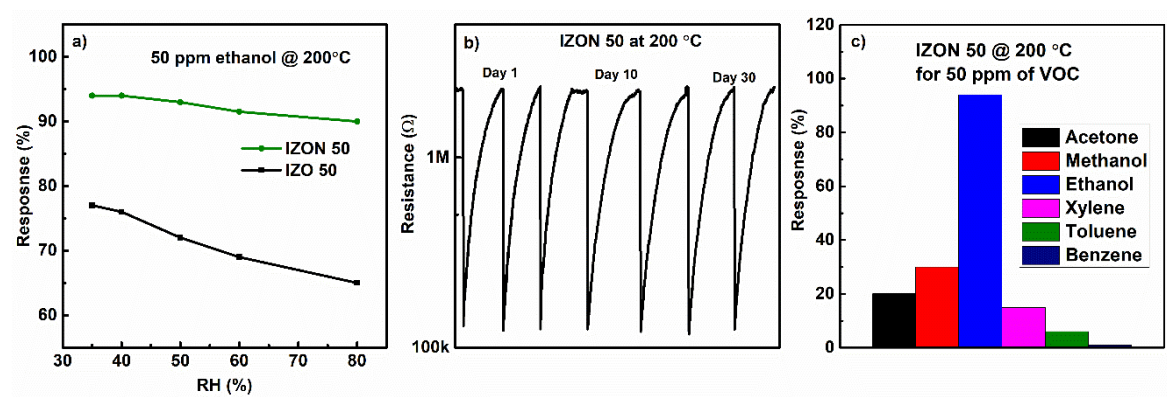

Figure 6 a) stability against relative humidity for both films, b) stability against time for IZON 50 thin film, c) selectivity of IZON 50 film for other VOCs at 200 °C.

Figure 6(a) compares the influence of relative humidity (RH%) on the ethanol sensing properties of IZON 50 and IZO 50. The IZON 50 displays significantly stable response (drop of only 5%) on the other hand the IZO 50 sensor suffers ~15 % reduction in the sensing response over the range of the RH from 10-80%. This observation agrees well with the reported literature that nitrogen incorporation inhibits water adsorption on metal oxides surfaces [22,30]. Reduced moisture adsorption leads to enhanced adsorption sites for oxygen, which results in higher ethanol sensing response even at high RH conditions[30]. Ethanol sensing response of IZON 50 has been examined over an extended period. As shown in figure 6(b) good base line recovery together with marginal variation in the sensor response has been attained over a period of one month. Figure 6(c) compares the selectivity of IZON 50 sensor for 50 ppm of ethanol, acetone, methanol, xylene, toluene and benzene measured at 200 °C. The composite film is selective to ethanol compared to other VOCs.

The superior ethanol sensing response of the film agrees well with earlier reported results on the ethanol sensing response of N-ZnO [25].

## 5. Mechanism

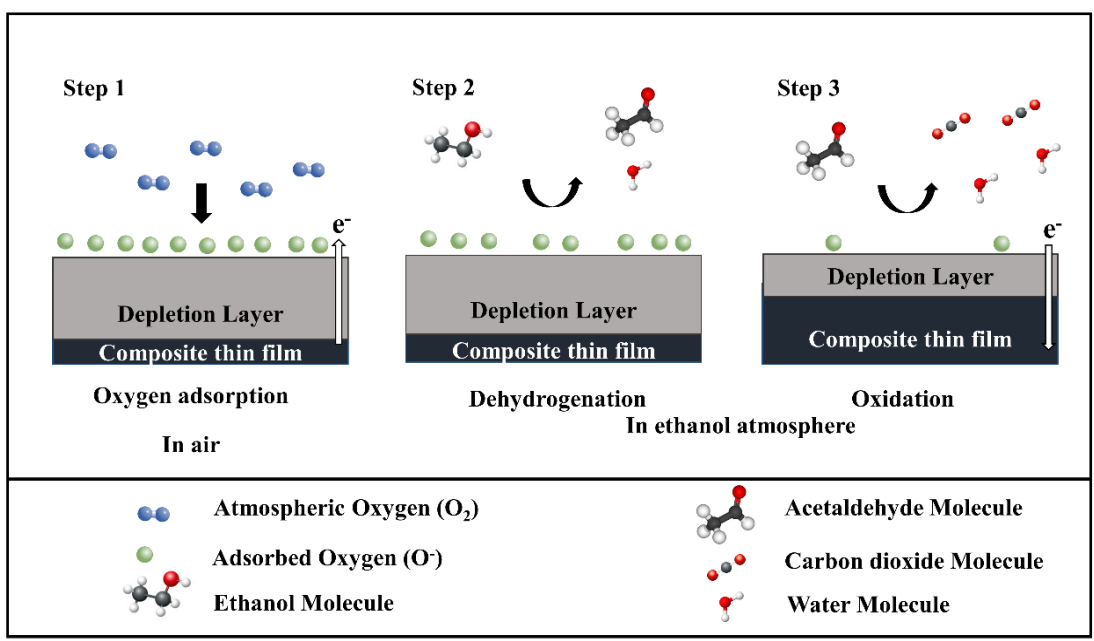

Figure 7 ethanol sensing mechanism of the composite thin film

In order to understand the improved ethanol sensing performance of IZON50, at first the ethanol sensing mechanism is considered. The sensing mechanism of composite thin films has been schematically represensted in figure 7. Similar to various n-type chemiresistive oxides, exposing the sensor surface to air leads to adsorption of atmospheric oxygen, which creates surface oxoanions (predominantly $O^-$ species at the sensor operating temperature of 200 $^\circ$C) [32] by capturing electrons from the conduction band of the oxide (step 1). This creates an electron depletion layer at the sensor surface which gives rise to potential barrier between individual grains and increases the overall resistance of the sensor. When the sensor is subjected to ethanol vapour, ethanol molecules get adsorbed on its surface and get oxidzed by the oxoanions (step 2 and step 3) to form carbon dioxide and water according to the reaction,

$$C_2H_5OH + 5O^- \rightarrow 2CO_2 + 2H_2O + 5e^- \quad (2)$$

This releases the electrons back to the conduction band of the oxides, which reduces the depletion layer width and thus the sensor resistance. As we had reported earlier, nitrogen doping increases the Vo cocentration in the chosen oxides that leads to an improvement in the dissociative adsorption of atmospheric oxygen on its surface [22]. Improvement of the receptor function of $In_2O_3$ (N-$In_2O_3$) and ZnO (N-ZnO) thin film based sensors because of nitrogen doping leads to their enhanced ethanol sensing performance in comparison to the undoped oxides [22,25]. Table S4 compares ethanol sensing response and operating temperature of N-$In_2O_3$, N-ZnO, IZO50 and IZON50 thin film sensors. Clearly the nitrogen doped nano-composite thin films exhibit superior response compared to the individual oxide thin films. Our observations agree well with the existing literature reports on the significant improvement of reducing gas sensing properties in $In_2O_3$/ZnO heterojunction based sensors [15,33]. It is to be noted that the ethanol sening performance has been significantly enhanced in case of the nitrogen doped nano-composite thin film sesnor (IZON50). The improvement of ethanol sensing performance in the nano-composite thin films has been primarily attributed to the formation of the nanoscale heterojunctions between $In_2O_3$ and ZnO.

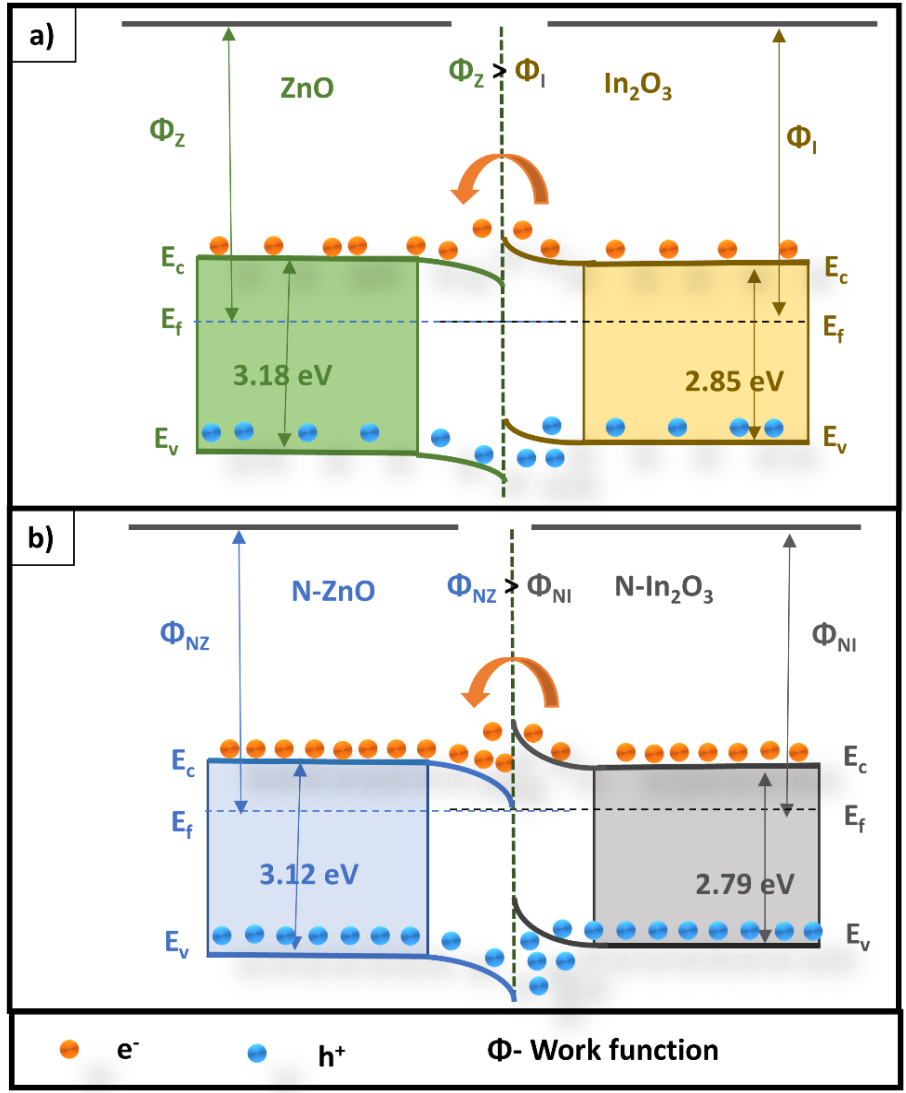

Fig 8. type-II heterojunction formation in a) IZO 50 and b) IZON 50 composite thin films

Formation of nano-scale heterojunctions with sharp interfaces between $In_2O_3$ and ZnO, and charge transfer from $In_2O_3$ to ZnO have been confirmed from TEM and XPS analyses respectively. Electron energy band alignment across $In_2O_3$ and ZnO interfaces has been reported to form type II heterojucntion [26]. Since the work function of ZnO is larger compared to $In_2O_3$, transfer of electron occurs from $In_2O_3$ to ZnO until the Fermi level aligns [24,34]. The band gaps of ZnO and $In_2O_3$ thin films have been estimated as 3.18 eV, 2.85 eV respectivly from diffuse reflectance spectroscopy study (figure S4) which agree well with those reported in the literature for oxygen deificent ZnO and $In_2O_3$ [35,36]. XPS investigation in the valence band region suggests separation between the valence band edge and Fermi level to be ~2.05 eV and ~1.95 eV for $In_2O_3$ and ZnO thin films respectively (figure S5(a&b)). Figure 8(a) schematically represents the charge transfer across $In_2O_3$/ZnO interface that leads to the formation of an electron accumulation layer on the ZnO side of the heterojunction. Electron accumulation at the interface gives rise to enhanced reaction sites for the adsorption of oxygen, which results in the improved sensitivity in the $In_2O_3$/ZnO nanocomposites [12]. It has been demonstrated that incorporation of nitrogen in the chosen n-type oxides increases their conductivity due to the enhanced concentration of the $V_{OS}$ [22,25]. The band gaps of N-ZnO and N-$In_2O_3$ thin films have been estimate as 3.12 eV and 2.79 eV (figure S4), while the separation between the valence band edge and Fermi level have been 2.2 eV and 2.18 respectively (figure S5( a&b)). This indicates stronger n-type nature of the nitrogen doped oxides compared to the undoped oxides which results in an enhanced electron density at the N-$In_2O_3$/N-ZnO interface in IZON 50 compared to IZO 50 as depicted schematically in figure 8(b). Enhanced electron concentration at the heterojunction interfaces in IZON 50 provides more reaction sites for oxygen chemisorption that leads to significant improvement in the sensor performance.

Superior performance of IZON 50 compared to the other N-doped nanocomposites is envisaged to be related to their microtructures. As discussed earlier, IZON 25 and IZON 75 possess agglomerated surface morphology, which reduces the active sites for gas adsorption [23]. On other hand, IZON 50 possesses mesoporous surface (SEM study) with large number of active sites. Larger surface pore coverage (40 %) in IZON 50 increases the active sites for adsorption. Table 1 compares ethanol sensing characteristics (e.g. response, operating temperature, response time and test gas concentration) of the present IZON50 sensor with those reported for various thin film and nanostructure-based sensors consisting of $In_2O_3$ and ZnO as the sensing materials. Inspecting the table, it is clear that ethanol sensing characteristics of IZON 50 thin film is superior compared to many of the routinely investigated sensors.

Table 1. Comparison of ethanol sensing characteristics of our study with literature

| Sl. No | Sensing Material | Working Temperature (°C) | Concentration (ppm) | Response (%) | Response Ra/Rg | Response time(s) |
|---|---|---|---|---|---|---|
| 1 | In-doped 3DOM ZnO [37] | 250 | 100 | - | 88 | N/A |
| 2 | Pd-ZnO nanorods [38] | 260 | 500 | - | 5.12 | 56 |
| 3 | N-$In_2O_3$ thin film [22] | 250 | 300 | - | 333 | 1 |
| 4 | N-ZnO thin film [25] | 225 | 300 | - | 324 | 12 |
| 5 | $In_2O_3$/ZnO nanostructure [14] | 250 | 300 | - | 900 | 25 |
| 6 | $In_2O_3$/ZnO core-shell nano fiber [17] | 225 | 200 | - | 57 | 4 |
| 7 | $In_2O_3$/ZnO nanosheet [15] | 240 | 100 | - | 300 | 45 |
| 8 | N-$In_2O_3$/N-ZnO composite thin film (this work) | 200 | 300 | ~99 | 324 | ~2 |

## 6. Conclusion

In summary, a robust, highly sensitive, selective, fast and reproducible ethanol sensor has been fabricated by wet chemical synthesis of nitrogen doped $In_2O_3$/ZnO nanocomposite mesoporous thin films on $SiO_2$/Si substrates. XPS investigation of the nano-composite thin films indicates nitrogen doping at the interstitial sites of $In_2O_3$ and substitutional sites of ZnO. TEM investigations confirm sharp interface between the $In_2O_3$ and ZnO grains indicating successful formation of n-n heterojunction. The uniqueness of the mesoporous nitrogen doped nanocomposite sensor with equal concentration of the oxide components is the availability of uniform pore distribution on the surface of the sensor and n-n hetero-junctions with higher electron concentration compared to undoped nano-composite sensor. This leads to significantly higher ethanol sensing response (∼94 %) and a fast response time (∼ 2 s) at a moderate operating temperature of ∼200 °C. The sensor exhibits excellent stability over a long duration and resilient to relative humidity conditions as high as 80%. Further, the sensor is highly selective to ethanol compared to various other volatile organic compounds. The experimental results suggest that incorporation of nitrogen in $In_2O_3$/ZnO nano-composite thin film is an effective route to improve the adsorption efficiency of ethanol molecules and facilitating the propagation of charge carriers across its surface. The present work presents a strategy to develop efficient ethanol sensor for utilization in law enforcement and biomedical applications both in terms of cost effectiveness and performance.